# Multi-Cloud Resource Provisioning with Aneka: A Unified and Integrated Utilisation of Microsoft Azure and Amazon EC2 Instances

Rajkumar Buyya[1,2] and Diana Barreto[1]

[1] Cloud Computing and Distributed Systems (CLOUDS) Laboratory
Department of Computing and Information Systems
The University of Melbourne, Australia

[2] Manjrasoft Pty Ltd, Melbourne, Australia

*Abstract*—**Many vendors are offering computing services on subscription basis via Infrastructure-as-a-Service (IaaS) model. Users can acquire resources from different providers and get the best of each of them to run their applications. However, deploying applications in multi-cloud environments is a complex task. Therefore, application platforms are needed to help developers to succeed. Aneka is one such platform that supports developers to program and deploy distributed applications in multi-cloud environments. It can be used to provision resources from different cloud providers and can be configured to request resources dynamically according to the needs of specific applications. This paper presents extensions incorporated in Aneka to support the deployment of applications in multi-cloud environments. The first extension shows the flexibility of Aneka architecture to add cloud providers. Specifically, we describe the addition of Microsoft Azure IaaS cloud provider. We also discuss the inclusion of public IPs to communicate resources located in different networks and the functionality of using PowerShell to automatize installation of Aneka on remote resources. We demonstrate how an application composed of independent tasks improves its total execution time when it is deployed in the multi-cloud environment created by Aneka using resources provisioned from Azure and EC2.**

## I. INTRODUCTION

The benefits of cloud computing have encouraged many prominent IT companies to invest in setting up cloud infrastructure. Therefore, there are an increasing number of cloud providers that are offering diverse products at the level of infrastructure, platform and software as a service [1].

Specifically, in the Infrastructure-as-a-Service (IaaS) model there are more than 50 providers offering computing resources [2] on subscription-basis using pay-as-you-go model. This presents a challenge in the selection of suitable cloud providers but at the same time, it opens an opportunity to create interconnected computing environments where the applications can provision resources from multiple cloud providers.

Running applications in cloud and multi-cloud environments brings important benefits. However, they should be developed to be able to exploit the main cloud characteristics. Therefore, software applications should have ability to acquire resources automatically without human intervention. Moreover, according to their current needs, they should add and release resources dynamically by measuring and monitoring usage of these resources [3].

Additionally, when applications are created to run on different clouds, it is important to consider other aspects such as transparent access to the services offered by different cloud providers, the migration of applications from one cloud to another, and the possibility of dynamically selecting the provider that best suits the need of an application [4, 5, 6].

Aneka is a Platform-as-a-Service (PaaS) that assists programmers to create applications that take advantage of cloud capabilities. It also helps developers to create applications using different distributed programming models. Additionally, it provides support to schedule and deploy tasks in cloud resources and offers services to dynamically acquire and release these resources from multiple clouds. Furthermore, Aneka components can be easily deployed in a provisioned virtual machine and supports monitoring the resources in which the applications are deployed [7].

Earlier version of Aneka had many features supporting multi-cloud computing. However, to provide wider support, additional functionalities are incorporated:

- Aneka used to support acquisition of resources from IaaS cloud providers using interfaces compatible with Amazon EC2, GoGrid and Xen and from the PaaS model in Windows Azure. Given that Azure changed its business model to allow acquisition of resources using the IaaS model, we extended Aneka functionality to support IaaS in Windows Azure.

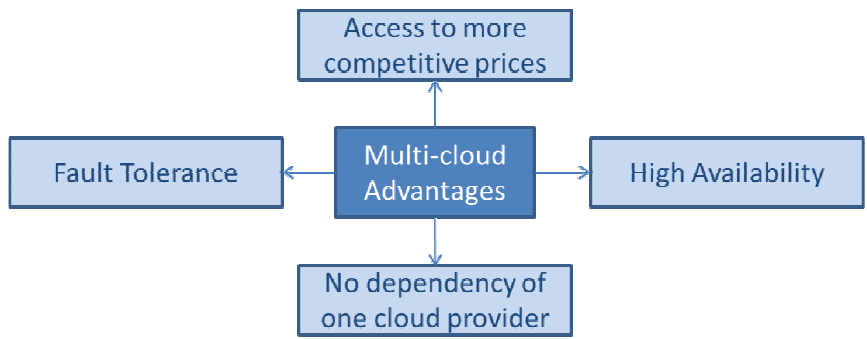

Figure 1. Advantages of multi-clouds.

- The previous version of Aneka was designed to install the Aneka container using the private IP by default, assuming that resources are located in the same network or virtual network. A new feature allows setting up Aneka to select between the public or private IP, so that machines from different networks can be connected via Internet.
- Finally, a new Aneka functionality allows automatic installation of Aneka containers in the machines provisioned dynamically from different cloud providers. Before, automatic installation existed for machines added statically using the Aneka Management Studio. To provide dynamic machines, it is necessary that users manually create an image with the Aneka installation in every cloud provider. Now a new installation can be done in every new dynamic provisioned machine and it is possible to include installations customized by the user using PowerShell.

The rest of the paper is organised as follows. Section II presents the advantages of multi-cloud environments and a general description of Aneka and its architecture. Section III presents the design aspects that are considered to extend the multi-cloud environment in Aneka. Section IV describes the improvements made in Aneka to provide more support to multi-cloud applications along with implementation details. Section V presents the results of experiments of running an application in a multi-cloud environment using Microsoft Windows Azure and Amazon EC2 cloud providers. Finally, Section VI concludes the paper with future directions.

II. BACKGROUND

*A. Multi-cloud Applications*

A multi-cloud platform is a middleware that allows interconnection of cloud environments by the use of an individual layer that deal with the issues of integrating and provisioning resources from different cloud providers [8]. Accordingly, Aneka can be categorized as a Multi-Cloud platform.

There are two different approaches for creating interconnected cloud computing environments. The first one is a *provider-centric* approach which is based on agreements among cloud providers. In this approach, a cloud provider can request other providers for more resources in order to satisfy the user requests and increase its profit. This leads to the creation of *cooperative federation of clouds*. The second approach is *client-centric* in which the client has the power of provisioning resources from multiple providers and can decide which provider to use to get the most benefit. Multi-cloud applications are part of this second group [8, 5].

Figure 1 summarizes the advantages of multi-cloud that are described as follows [4, 5, 6]:

- **Access to more competitive prices:** Users can select and compare prices that providers offer for different configurations. Then, they can choose the best configuration for their needs and the best price in the market for a specific configuration.
- **High availability and improved response time:** When resources are distributed in different clouds, having the applications closer to the user improves the time of response and the applications availability.
- **Fault tolerance and reliability:** If the applications are replicated in different cloud providers and different geographical regions, the risk of interruptions due to infrastructure failures could be minimal, because if one provider fails, resources from other cloud provider will be able to keep on responding user requests.
- **No dependency of one cloud provider:** Applications that can be run in different cloud providers should be easily migrated, therefore if the user does not want to continue working with a specific provider, he can choose and move to another one transparently.
- **Simplify the integration of on premise and cloud resources:** Applications that cannot be totally migrated to cloud environments can use multi-cloud platforms to provide the environment and support for a successful integration with on premise resources.

B. *Aneka: A Cloud Application Platform*

Aneka is a Platform-as-a-Service (PaaS) that facilitates the development, deployment and monitoring of distributed applications in clouds. For this purpose, Aneka has different components that are described below [7]:

- **Aneka Client Libraries:** These are Application Programming Interfaces (APIs) created to build applications based on different distributed programming models including Task, Thread and Map-Reduce models. After developing an application using these APIs, it can be easily deployed and executed in Aneka managed clouds.
- **Aneka Cloud:** This is the virtual network that is composed by the Aneka Master and the Aneka Worker Containers which work together to run the applications. The resources used for the Aneka cloud can be provisioned from different sources such as private and public clouds, networks of computers, or multicore servers.
- **Aneka Containers:** These are the Aneka components that are deployed on different machines that are part of the Aneka cloud. These containers host and offer different services that vary according to the two roles that can be installed: Aneka Master and Aneka Worker. An important feature of Aneka containers is that they are developed using a Service Oriented Architecture (SOA), so containers are composed of a core and a number of services that can be easily plugged and unplugged according to the needs of the user applications. Furthermore, it is possible to create new services or replace the existing ones. This makes Aneka a highly customizable and extensible system.
- **Aneka Master:** This is an Aneka container whose main responsibility is orchestrating and monitoring the execution of the application created by the user. Therefore, its default services are scheduling, provisioning, storage, reporting and accounting.
- **Aneka Worker:** This is an Aneka container that is responsible for executing every individual task that constitutes the user application. The default services associated to this role are allocation and execution.
- **Aneka Management Studio:** This is a graphical interface that allows the construction and monitoring of the Aneka cloud. Static and dynamic Aneka clouds can be configured. In the static cloud, the required resources are added manually while in the dynamic cloud an Aneka Master is set up with a provisioning service to request and release machines dynamically from cloud providers.

The monitoring capabilities of Aneka are supported via the reporting service, the container core and the Aneka Management Studio. The Aneka Management Studio acts as the central point that communicates with the containers and its services to generate reports to observe the status of the containers, resource utilization details and associated cost of the resources used.

Figure 2 shows a general view of the Aneka components. It shows an Aneka cloud composed of one Aneka Master and six Aneka Workers that have been provisioned from private and public clouds. These resources can be added to the cloud in a static or dynamic way. Also it is possible to observe an application that has been developed using the Aneka client libraries and run in a distributed way using the Aneka cloud resources. Finally, it shows the Aneka Management Studio as a key tool that allows creation and monitoring of the Aneka cloud.

III. ANEKA MULTI-CLOUD DESIGN

A. *General Design*

As noted earlier, Aneka is composed of services that can be plugged and unplugged. In this section, we focus on the Aneka services that support some of the suggested multi-cloud requirements [5]. Figure 3 shows these services, the Aneka components where they are contained and it also highlights the services that are modified to extend the Aneka multi-cloud environment. A brief description of these services is provided below:

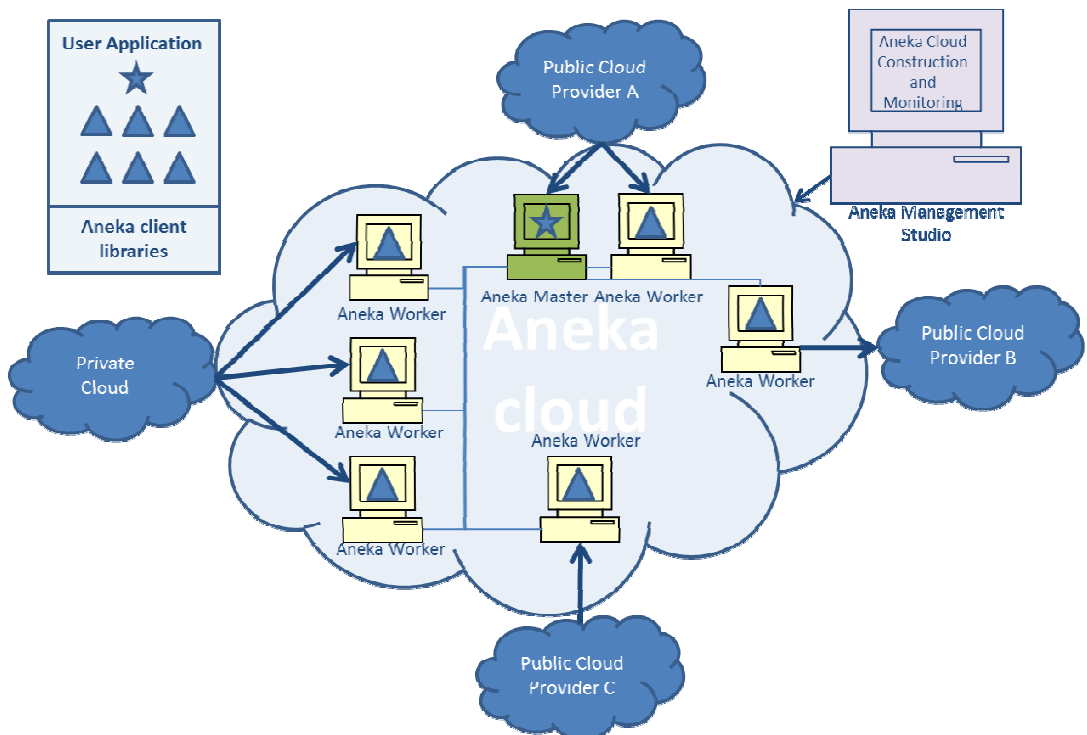

Figure 2. Aneka components.

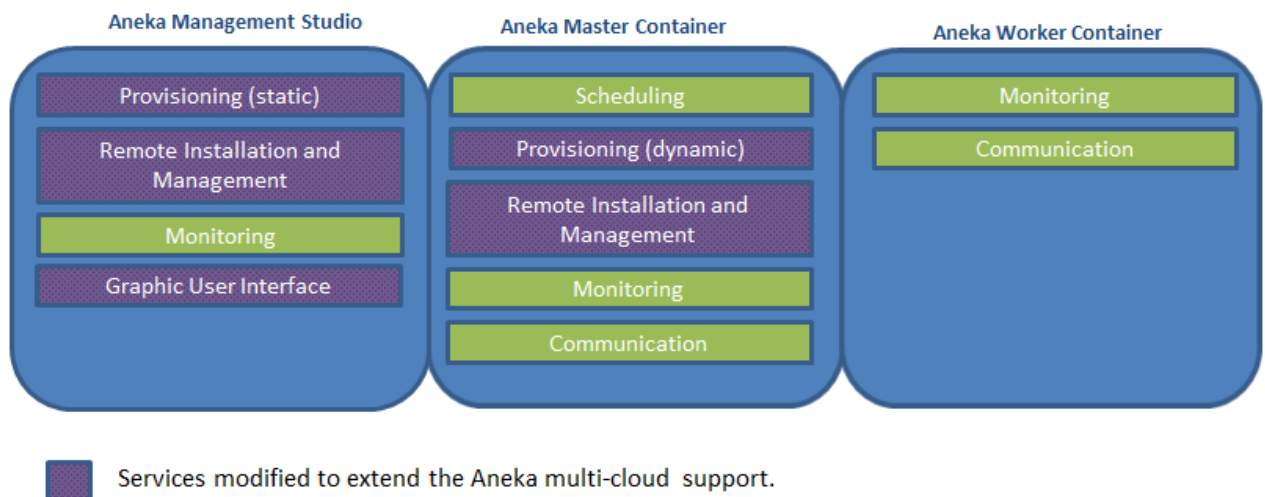

Figure 3. Multi-cloud Aneka.

- **Scheduling:** The scheduling service is in charge of assigning tasks to available resources. In a static configuration, this service uses scheduling algorithms to assign tasks to resources that are previously provisioned. However, in a dynamic configuration, the user should configure scheduling algorithms that decide when to provision or release resources directly from cloud providers. Note that these tasks are performed through the provisioning service.

The scheduling algorithm is an important component to achieve a successful deployment of distributed applications in multi-cloud environments. Scheduling algorithms can be developed to fulfil Service Level Agreements (SLA) required by the customers. For example, Aneka uses a deadline-driven algorithm to request the resources that are needed to finish the applications tasks before a specified time [9].

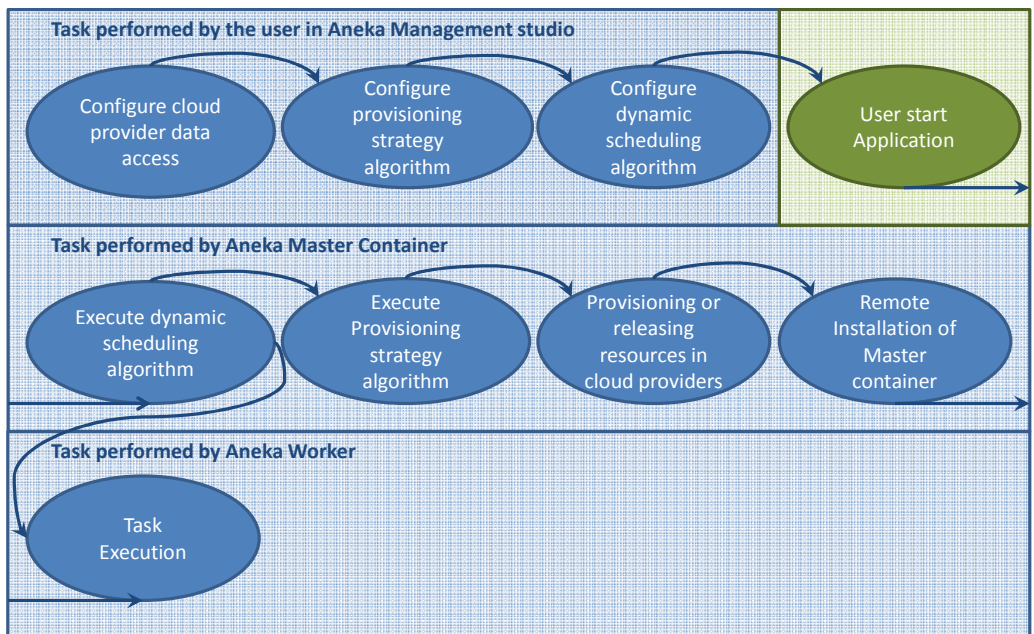

Figure 4. Complete resource provisioning process.

- **Provisioning:** The main functionality of the provisioning service is to provision and release resources that are acquired from different cloud providers. To satisfy this requirement, Aneka offers abstract interfaces that can be used to implement the logic to invoke cloud provider interfaces to provision and release resources.

  Additionally, Aneka provides standard editors to configure the access data to the cloud providers and the virtual machines features. These configurations are used to create a pool of connections that can be invoked any time such that resources from a specific cloud provider can be acquired or released.

  In addition to the pool of connections to cloud providers these service also needs to be configured with an algorithm that specify the strategy to select the cloud provider that should be used to acquire the next resource to be provisioned. This strategy is invoked by scheduling algorithms when the resources are acquired dynamically by the Aneka Master Container. Figure 4 shows the process of the dynamic provisioning of resources in Aneka.

  When the provisioning service is invoked by the Aneka Management Studio, the resources are added manually and statically by the user. In this case, the user specifies the cloud provider that he wants to use to provision the virtual machine and the provisioning strategy algorithm is not necessary.

- **Remote Installation and Management:** This service is created to install and manage Aneka containers and user applications after provisioning virtual machines. Therefore, the duties of this service are downloading executables and configuration files of Aneka, and user applications from a remote repository, installing the containers services and user application on the remote machines, stopping and starting services, and uninstalling containers services and user applications.

  For installation, Aneka offers two approaches. The first one consists of automatically downloading the installers and configurations files from a repository to perform a complete installation in each new provisioned machine. Using this approach the installation is totally independent of the cloud provider. However, a more efficient deployment is achieved when preconfigured images are used. Therefore, the second option is to use Aneka installation service to generate a base image for each cloud provider. Then Aneka will use these images when the machines are provisioned and the installation service is used just to configure dynamic aspects such as IP addresses.
- **Monitoring:** The monitoring service happens in three Aneka components. Aneka Workers Containers have the responsibility of monitoring the execution of the allocated task and the consumption of their own resources. On the other hand, Aneka Master Containers monitor the execution of the whole application and report this information to the scheduling service that acts according to the configured algorithm. Finally, the Aneka Management Studio monitors the individual containers and generates reports to the users for visualizing the performance of the complete Aneka cloud.
- **Communication:** Aneka allows communication among worker and master containers using the TCP protocol, so that if the provisioned machines are part of the same network or virtual private network (VPN), they can be configured to communicate using the private IP. However, it is not always the case and sometimes it is necessary to communicate machines in different networks via Internet using their public IP.

TABLE I
MULTI-CLOUD ANEKA SERVICES DESIGN REQUIREMENTS.

| **Scheduling** | **Provisioning** | **Remote Installation and Management** | **Monitoring** | **Communication** |
|---|---|---|---|---|
| Configuration of QoS policies. | Abstract interfaces to add new cloud providers. | Customized installation of services application. | Profiling the consumption of resources of virtual machines. | Communication among machines provisioned from different cloud providers. |
| Request resources according to QoS policies. | Cloud providers access data repository. | Installation of Aneka Container and user application independent of the cloud provider. | | |
| Reallocation of failed tasks in resources of current available cloud providers. | Cloud provider selection logic. | | | |
| | Functionality to provision and release resources. | | | |

The Aneka extensions are mainly located in the services of Provisioning and Remote Installation Management. It is important to note that the Aneka Management Studio is the user interface to interact with the whole Aneka infrastructure and users can use it to configure services. Therefore, in order to modify the services, it is necessary to adjust it. Additionally, changes in the configuration of master and workers container are added to allow communication via public IP.

Table I summarizes the services and the multi-cloud design requirements [5] that are addressed for each of them. In order to show the flexibility of Aneka to work in multi-cloud environments, we now discuss software design details of Aneka services where the main changes are made.

*B. Provisioning Service*

Provisioning service is responsible for the connection with different cloud providers and for the selection of the provider that should be used to acquire the next resource to be added to the Aneka cloud. Therefore, it is an important service for deploying and executing applications in multi-cloud environments. Aneka has been designed to facilitate the inclusion of new functionality to this service as shown in the class diagram in Figure 5, which shows an entry point called ResourceProvisioningService. This class receives an external message for provisioning or releasing virtual machines from cloud providers. To do its work, the class invokes ResourcePoolManager. It uses the available IResourcePools and the IProvisioningStrategy to fulfil the tasks. The IResourcePool has the logic for provisioning and releasing resources for a specific cloud provider. The IProvisioningStrategy is the class that has the algorithm to select the IResourcePool that will be used to request a new virtual machine.

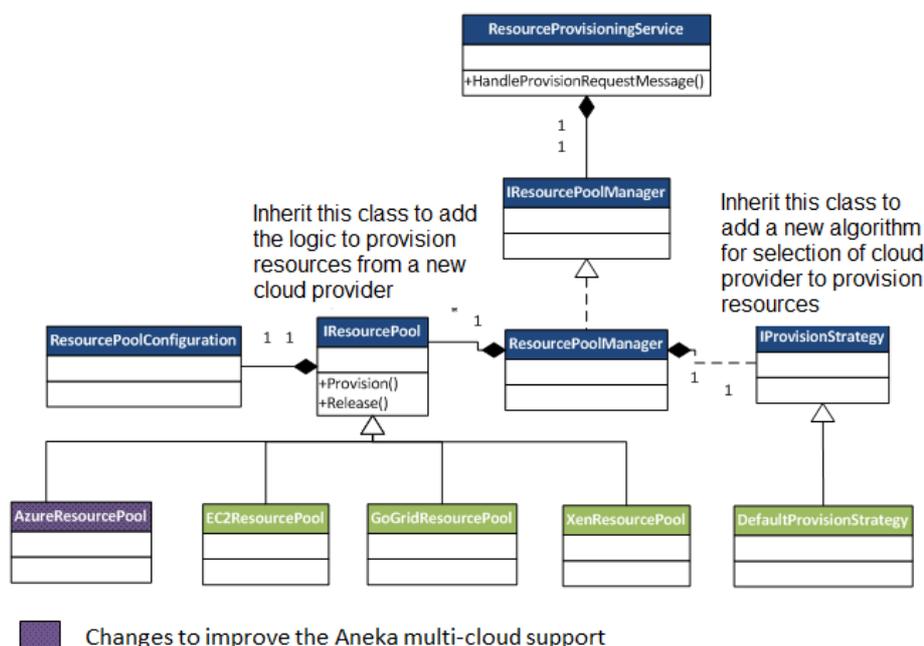

Figure 5. Provisioning service class diagram.

In order to include a new cloud provider for provisioning resources, an Aneka user can create a new class that inherits from IResourcePool. The implementation of a class that extends from IResourcePool should be able to communicate with the API of the cloud provider to provision or release resources. For each IResourcePool, there is an IResourcePoolConfiguration class that should be instantiated using the access data to the cloud provider and additional features of the resources to be requested. For instance, if the IResourcePoolConfiguration belongs to the EC2ResourcePool it should contain the access key and the secret access key which are the parameters used to verify the Amazon account has access to request resources. It should also contain information such as the type of virtual machines that will be requested.

New cloud providers' resource-pools are easily added by inheriting the class IResourcePool. Similarly, new strategies to choose and prioritize IResourcePool can be added by inheriting IProvisionStrategy. In the implementation of the provisioning strategy, users can implement a simple algorithm that prioritizes the IResourcePools using a priority value given by the user. They can also use more complex algorithms, for example, external services such as cloud harmony [2] to know which cloud server is having the best performance.

In summary, the provisioning service is the one that makes possible a dynamic and wise acquisition of multi-cloud virtual machines. The next section describes more details about the updates made to Aneka specifically in the context of remote installation and management service.

### C. Remote Installation and Management

After resources are provisioned, it is necessary to install Aneka Worker Container in the machines to enable Aneka Master to recognize and manage these new resources. The remote installation and management utility service is designed to install and manage the installation of Containers in the machines that form the Aneka cloud. Figure 6 shows the class diagram of this service.

The functionality of this service is accessed by the class DaemonManager, which receives the platform details (Windows or Linux) for accessing the machine and it uses the right classes to perform the remote operations. In Figure 6, we note some examples of possible platforms. The classes related with Windows and Linux platform are used to perform remote operations in machines running respective operating systems located in the same network (or virtual network) of the master machines, while Windows Public is defined to add and manage Windows machines that are located in different networks. This also can be used to perform installations customized by the user.

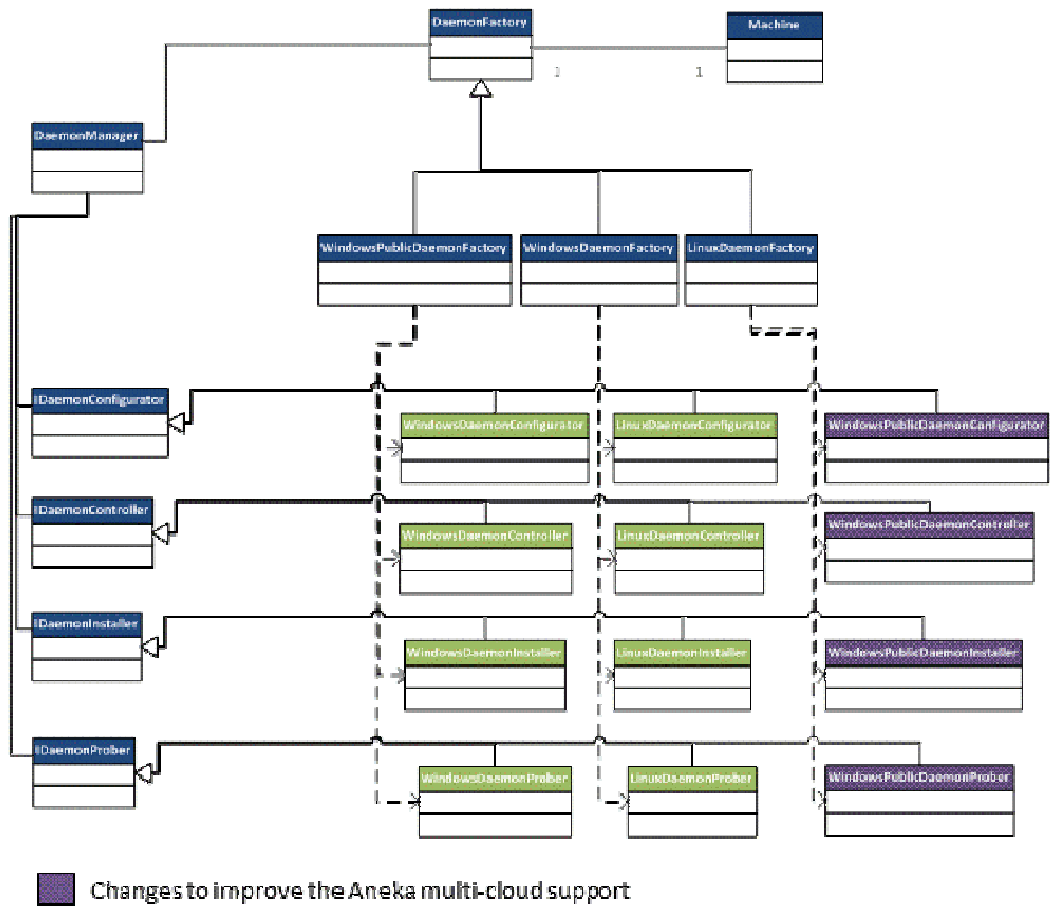

Figure 6. Installation and management service class diagram.

The base classes and interfaces are discussed below:

- **IDaemonConfigurator:** The classes that inherit this interface are used to load the default configuration of the Aneka installation. For example the home directory where the Aneka container should be installed is loaded there.
- **IDaemonProber:** The classes that inherit this interface are used to probe the status of the remote Aneka container installation.
- **IDaemonController:** The classes that inherit this interface are used to control the remote Aneka service. Using this it is possible to start, stop, and restart the Aneka service in the remote machine.
- **DaemonInstaller:** This class is used to install and uninstall the Aneka service.

IV. EXTENDING ANEKA MULTI-CLOUD ENVIRONMENT

A. *Adding Microsoft Azure IaaS as Aneka Cloud Provider*

It is important for a multi-cloud platform to have access to the best recognized Cloud Providers [5] and Microsoft Azure is one them. Previously, Aneka offered the functionality of deploying containers in the Microsoft Azure Platform-as-a-Service (PaaS). However, Microsoft Azure changed the business strategy to offer Infrastructure-as-a-Service (IaaS) model. Therefore, it is necessary for us to create a new

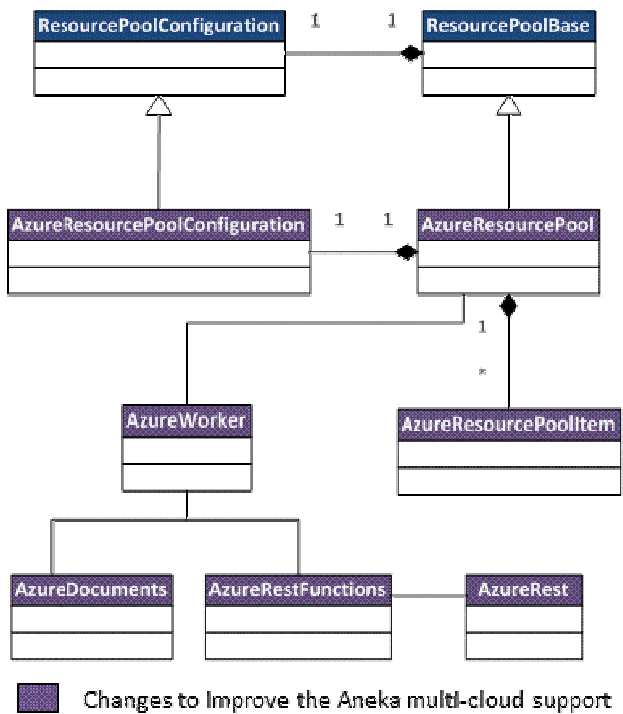

Figure 7. Azure ResourcePool class diagram.

IResourcePool in Aneka to enable acquisition of virtual machines using the new APIs provided for Azure.

Microsoft Azure allows customers to access their IaaS services through REST APIs [10]. As a result, most of the work that is developed in Aneka to access Azure is focused on creating the XML documents to send the information that describes the required resources for invoking Azure APIs. Figure 7 shows the class diagram with the classes that are added to include Microsoft Azure to provision resources. Then these classes are explained:

- **AzureResourcePool:** This class is the entry point to access Microsoft Azure resources. It inherits from the class IResourcePool and in this way it is transparently included in the Aneka design. To implement the IResourcePool methods, Provision() and Release(), it keeps a reference to the class AzureWorker. To maintain the user configuration, it uses an instance of the class AzureResourcePoolConfiguration.
- **AzureWorker:** This class has the logic to orchestrate the operations that are required by Microsoft Azure to request resources. To fulfil its duties, this class has an instance of the class AzureRestFunctions that has the logic to access the specific functions of the Microsoft Azure REST API.

In order to acquire a new virtual machine in Microsoft Azure, it is required to associate it to a cloud service and storage. A cloud service is a virtual private cloud that allows the machines inside it to communicate among each other and use the same public IP. The storage is the place where the virtual machine data is kept. The user can decide if he wants to reuse a specific cloud service and storage or if new ones are required. Similarly, a user can decide if he wants to use public or private IP and other details about the machines required. The logic to request resources according to user requirements is located in AzureWorker class.

- **AzureRestFunctions:** This class uses AzureDocuments to generate the XML documents that are required for the rest functions. Additionally, it uses AzureRest for the final rest communication with Windows Azure. Using these two classes, AzureRestFunctions create the functionality to request, query and remove the specific resources needed for provision resources in Aneka. Therefore, this class has the method to create and release storages, cloud services and virtual machines.
- **AzureDocuments:** This class generates the XML documents that are sent to the Aneka Rest web-services to specify the features of the resources required.
- **AzureRest:** This class has the basic methods (get, post and delete) to access the REST APIs of Microsoft Azure.

### B. Supporting Communication Via Public IP

For communication between the Aneka Master and Worker Container, Aneka assumes that all the machines belong to the same network. Consequently, the private IP of the machine where Container is installed is used for communication. This works when the resources are located in the same network. This is not always the case as resources can come from outside networks. Although a Virtual Private Network (VPN) solves this issue, there are cases users want to access resources from different networks without configuring a VPN.

To support communication via public IP, Aneka allows users to configure the parameters to provision machines from cloud providers' network type: private or hybrid. Using this information, the machines provisioned are able to configure the Aneka containers to use the public or private IP. To obtain the public IP of specific machine, Aneka uses a PowerShell script that is installed together with the Aneka

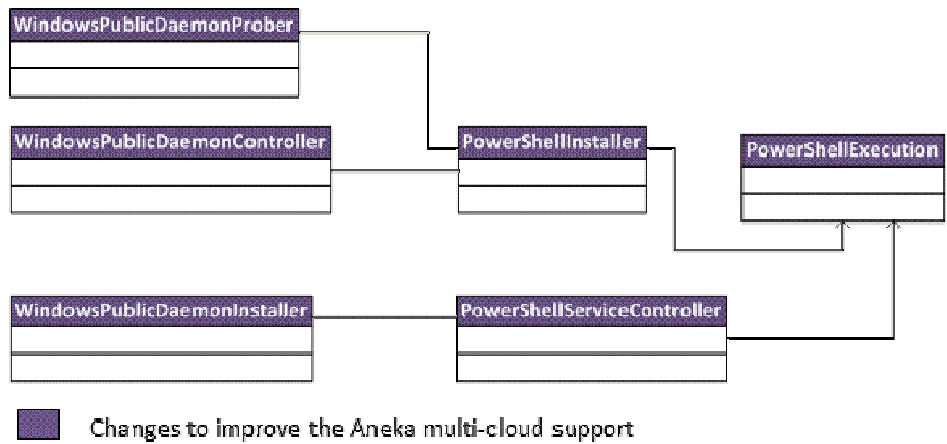

Figure 8. Classes to dynamic installation of Aneka components using PowerShell.

Container. Knowing the public IP address and opening the communication ports, it is possible for Master and Worker Containers located in different networks to communicate.

*C. Improving the Automatic Installation of Aneka Containers*

Aneka supports automatic installation of Aneka containers in machines that belong to the same network and are added statically via the Aneka Management Studio. However, when machines are added dynamically from cloud providers an installation is performed. Therefore, it required that machines are provisioned using an image that has the Aneka Worker Container and the environment configuration previously installed. This way of installation requires creation of a new image every time different configuration of Aneka is needed from a cloud provider.

The first modification, done in Aneka to avoid creating new images frequently, is to invoke the remote installation and management service from the provisioning service. In this way, after the machines are created, the required components are installed. However, installation of all Aneka components is a heavy task (time consuming), which makes the provisioning process slower. Therefore, a new set of installation classes are designed to support customized installations. It means that images with preinstalled Aneka components should be used, but the user can provide instructions, for example, to change parameters such as the Aneka Master IP and to run small software installations.

The new classes shown in Figure 8 are developed using PowerShell. This technology has the advantage not just allowing users to customize their installation scripts, but also to install and manage remote windows servers using the public IP. The PowerShellInstaller class has the logic to run the commands to install and uninstall the Aneka Containers while the PowerShellServiceController class contains the logic to start and stop the windows services associated to the Aneka Containers. Finally, PowerShellExecution has the logic to execute any set of PowerShell commands.

*D. Technical Details – Implementation Technologies*

Aneka platform is developed using .NET technologies including C# programming language and the .NET framework (version 3.5). The Aneka Management studio is a desktop application developed using the default .NET components and the Aneka Worker and Master Containers are deployed in the provisioned machines using Windows NT services.

The files to install Aneka Containers and user applications are located in remote repositories. If the services are all located in the same Virtual Private Network (VPN) and the machines are Windows, it is possible to use Windows File Sharing repository. For access this repository, Aneka uses WNet API that allows to access resources inside of the same network. If the machines are not located in the same network, Aneka supports the use of FTP repositories.

To automate the installation and configuration process of the machines that are provisioned from the cloud provider, a key technology we use in Aneka is Windows PowerShell. This tool provides system administrators with command line and scripting language that allow tasks to be run both in local machines and in remote machines [11].

The virtual machines are instantiated by enabling remote PowerShell and using PowerShell script to open ports, configure IP address in configurations files, and perform the installation process.

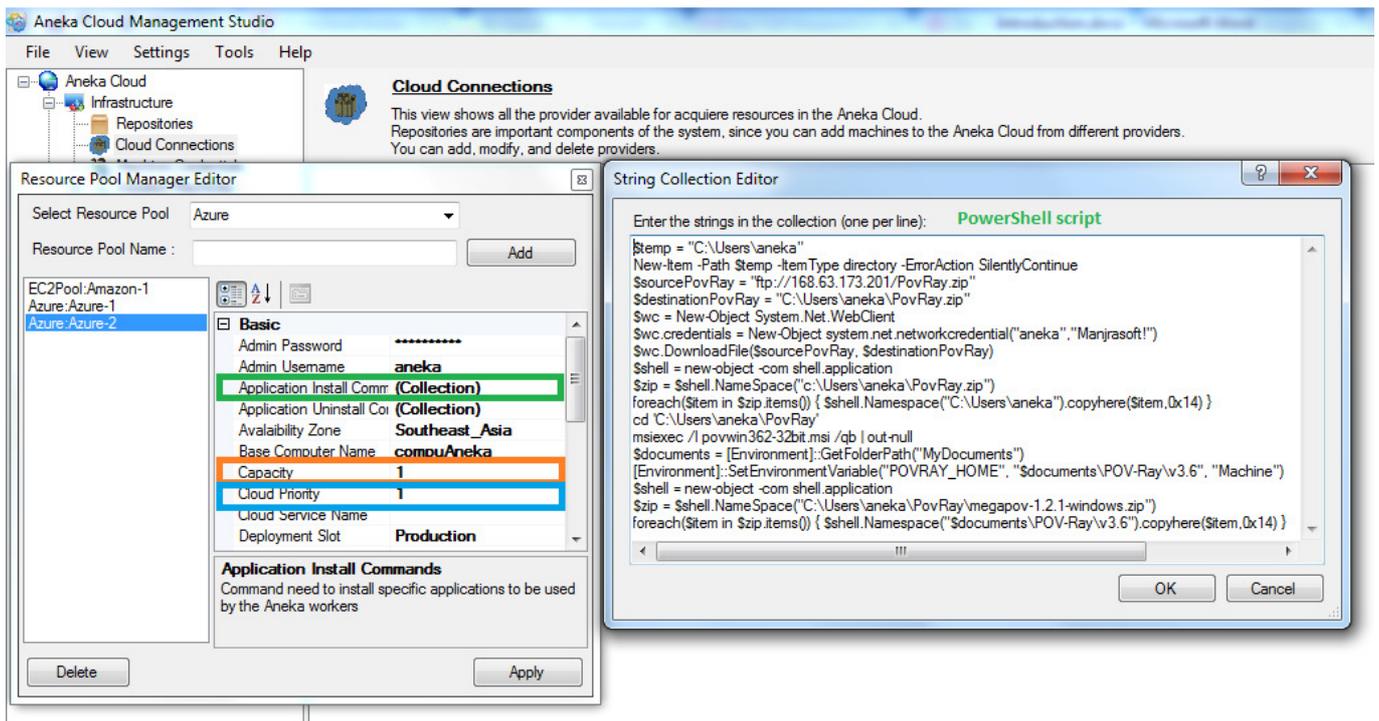

Figure 9. Cloud providers' configuration.

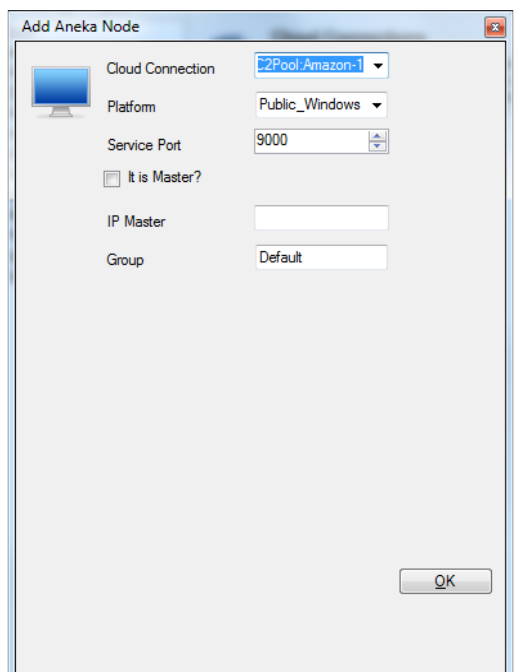

Figure 10. Screen shot for adding statically a machine from a specific cloud provider.

### E. Multi-cloud Environment Configuration

As noted earlier, the Aneka Management Studio offers capabilities for configuring the multi-cloud computing environment for integrating resources from different providers. In this section, we describe the configuration screen shots to highlight tools that Aneka provides for creating a flexible multi-cloud environment.

First, in order to create static multi-clouds, Aneka permits saving of the configuration to connect to several cloud providers and more than one connection can be created for an individual provider. Together with the data connections required by access resources in the cloud provider, it is possible to provide a PowerShell script that is executed when Aneka is installed in the machine. Additionally, user can also configure the maximum number of machines that can be provisioned and a priority number. This priority number is an input for the provisioning strategy algorithms. Figure 9 shows the Aneka screen shot that asks for the fields described.

After creating the connection, the user can create a static multi-cloud environment to add the machines manually. Figure 10 illustrates the screen shot that is shown to the users when they select the option to add machine. To create a new static machine, the user should inform if this machine should be installed as Aneka Master; otherwise, it should provide the IP of the Aneka Master so that the new Aneka Worker can communicate with the Master. Users should also provide the number of the port that they want to use to install Aneka, the platform, and the name of group associated to this Aneka Cloud. User can create as many machines as it specified in the Capacity field of the connection configuration.

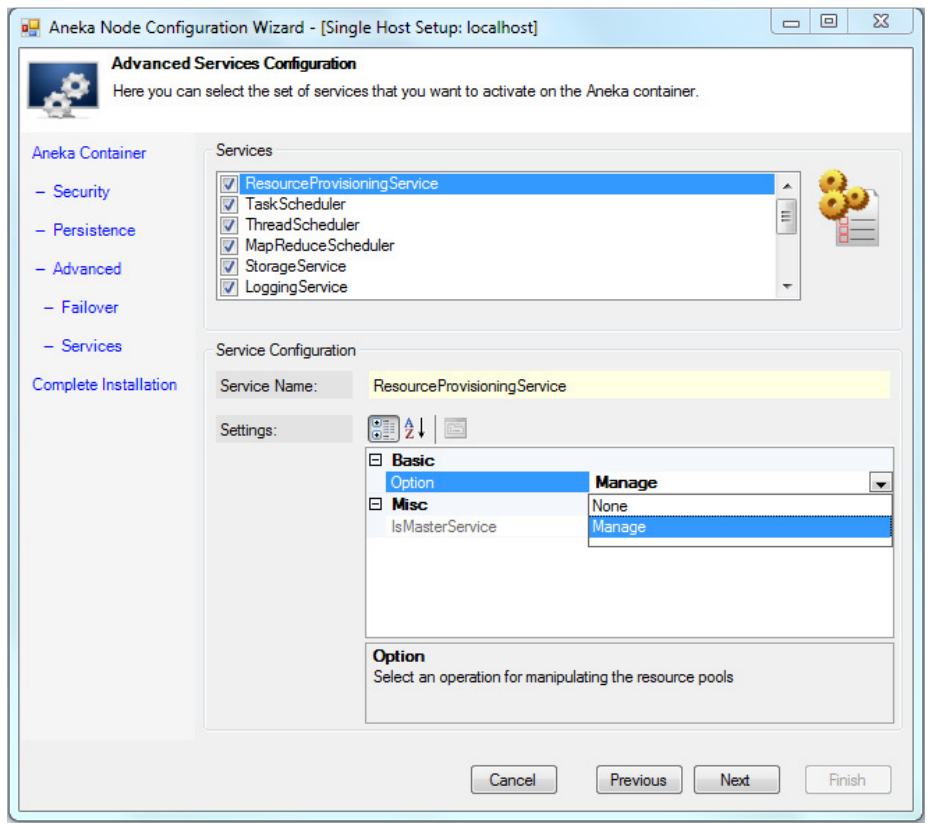

Figure 11. Screen shot for adding statically a machine from a specific cloud provider.

After the machine is provisioned, the user can give the instruction to install Aneka and run the user script in the remote machines. Then any application created with Aneka can run on this static Aneka cloud.

Creating a static cloud is useful if the user has constant access to a fixed number of resources or if the application needs a fixed set of resources. Otherwise, a more interesting approach is to create dynamic clouds. A user can configure a dynamic cloud when he configures the Aneka Master Container. Figure 11 shows the screen shot illustrating configuration of services. The user should indicate the Provisioning service that he wants to use dynamic configuration by selecting the option "manage provisioning". Currently, Aneka supports one algorithm related with the provisioning strategy. If multiple algorithms are supported, from this screen shot, the user should be able to select the suitable algorithm.

In this service configuration, the user should also select the scheduling algorithm that the programming model uses to request and release machines. Figure 12 shows the available scheduling algorithms for the task programming model. The two supported algorithms for dynamic provisioning are:
- DeadlinePriorityProvisioningAlgorithm and
- FixedQueueProvisioningAlgorithm.

However, new algorithms can be developed and plugged by Aneka developers – an example demonstrated by researchers from Washington State University, USA [14].

After the user configures and installs the Aneka Master Container, it is able to automatically provision, install, and release virtual machines from different cloud providers. The complete multi-cloud features of Aneka are available for the cloud providers, Amazon EC2 and Windows Azure. However, Aneka architecture supports creation of new plugins for integrating other cloud providers.

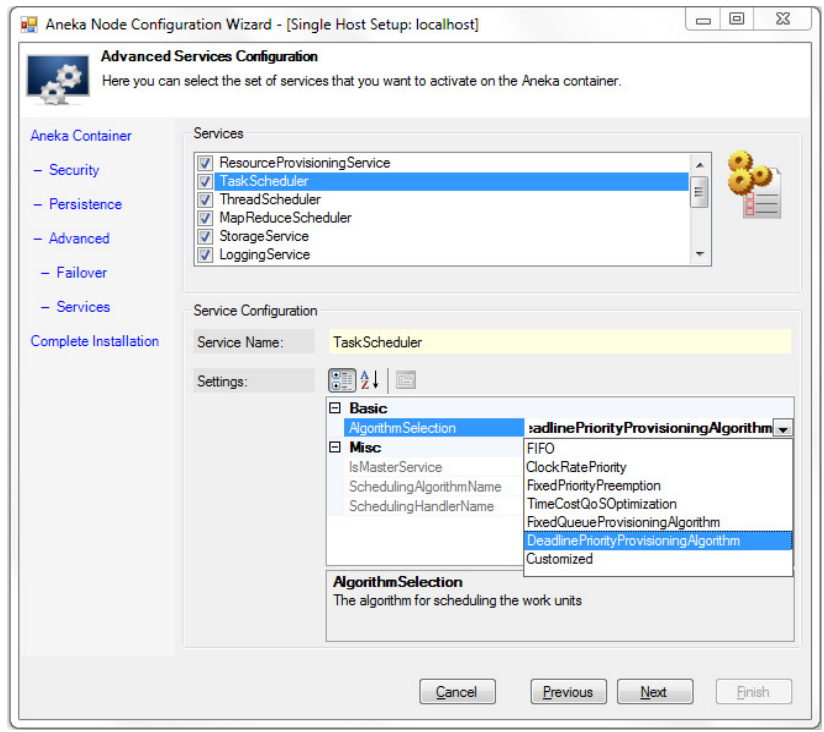

Figure 12. Screen for adding a machine statically from a specific cloud provider.

## V. PERFORMANCE EVALUATION

To evaluate the functionality of Aneka in multi-cloud environments, we run an application that uses the task programming model for composing a bag-of-tasks or parameter sweep application. The application executed is called BLAST (Basic Local Alignment Search Tool) [13]. BLAST is tool for searching sequences of nucleotide or protein in one or several databases. We downloaded a database from the NIH National Centre for Biotechnology Information (NBCI), USA. We used the database "ecoli.nt" whose size is of 4652 kb and each task executes one query, so we are able to perform multiple queries over the same database. Each query has a size of 1kb.

We created a static Aneka cloud composed of machines provisioned statically from Amazon EC2 and Microsoft Azure with similar configurations and locations. The Amazon EC2 virtual machines are located in the Asia Pacific (Sydney) datacenter and running Windows Server 2012 R2 Base of 64 bits and are type small instance with configuration of 1 virtual CPU and 2 GB of memory. Similarly, Microsoft Azure virtual machines are located in Southeast Asia datacenter, their operating system is Windows Server 2012 of 64 bits and the type is small instance with configuration of 1.75 GB of memory and 1 CPU core.

Table II shows the results of the experiments. The number of machines used to run the tasks was incremented gradually showing how the total execution time was reduced when more machines were added to the Aneka cloud. With one Aneka worker running on Azure instance, execution of 80 tasks of BLAST application finished in 15 minutes and 53 seconds. With 6 Aneka workers (3 running on Azure and 3 on EC2), the application took 1 minute and 50 seconds.

We note that the execution time is reduced with provisioning additional resources regardless of their provider. We conclude that for application with independent tasks that require low communication, the use of multiple cloud providers does not have significant impact on the performance.

## VI. CONCLUSIONS AND FUTURE WORK

We presented the benefits of multi-cloud computing: access to competitive prices, high availability, fault redundancy and reliability, no dependency on a single cloud provider. To simplify the integration of resources from multiple cloud providers, we introduced the Aneka multi-cloud platform and its components that make it possible to run distributed applications in multi-cloud environments. In this context, the two Aneka services described in detail are: provisioning and remote installation and management.

TABLE II
TOTAL EXECUTION TIME OF A SET OF TASKS DISTRIBUTED IN TWO CLOUD PROVIDERS.

| Number of Tasks | Virtual Machines/Resources | | Total Workers | Duration (HH:MM:SS) |
|---|---|---|---|---|
| | Microsoft Azure | Amazon EC2 | | |
| 80 | 1 | 0 | 1 | 0:15:53 |
| 80 | 2 | 0 | 2 | 0:07:53 |
| 80 | 3 | 0 | 3 | 0:05:28 |
| 80 | 3 | 1 | 4 | 0:04:30 |
| 80 | 3 | 2 | 5 | 0:03:12 |
| 80 | 3 | 3 | 6 | 0:01:50 |
| 160 | 1 | 0 | 1 | 0:34:08 |
| 160 | 2 | 0 | 2 | 0:20:44 |
| 160 | 3 | 0 | 3 | 0:13:06 |
| 160 | 3 | 1 | 4 | 0:09:20 |
| 160 | 3 | 2 | 5 | 0:07:21 |
| 160 | 3 | 3 | 6 | 0:03:36 |

We introduced the new extensions made to Aneka including how Aneka provisioning service is modified to support Microsoft Azure IaaS and communication via public IP. We also described how the remote installation and management service is modified to improve the automatic installation of Aneka containers using PowerShell.

Using the Aneka platform, we carried out experiments to demonstrate execution performance of the BLAST application running in the multi-cloud environment.

In order to avoid the code modification every time that providers change their interfaces, as a future work, Aneka can be extended to use features of JClouds [12]. New algorithms for selection of cloud providers based on user requirements such as privacy, market models, energy-efficiency, and other QoS parameters can be implemented. It can be further extended to support provisioning of not just virtual machine resources, but other resources such as storage resources and software-defined networks as well.

**Acknowledgements**: We thank Raghavendra Kune, Rekha Mangal, Rajinder Sandhu, Rodrigo Calheiros, Amir Vahid, Chenhao Qu, Adel Nadjaran Toosi, and Xinghui Zhao for their comments on improving the work.